# Nonlinear Nanomechanical Mass Spectrometry at the Single-Nanoparticle Level

*Mert Yuksel[†], Ezgi Orhan[†], Cenk Yanik[‡], Atakan B. Ari[†], Alper Demir[#], M. Selim Hanay[†,§,*]*

[†] Department of Mechanical Engineering, Bilkent University, 06800, Ankara, Turkey

[‡] Sabanci University SUNUM Nanotechnology Research Center, 34956, Istanbul, Turkey

[§] National Nanotechnology Research Center (UNAM), Bilkent University, 06800, Ankara, Turkey

[#] Department of Electrical Engineering, Koc University, 34450, Istanbul, Turkey

ABSTRACT

Nanoelectromechanical Systems (NEMS) have emerged as a promising technology for performing the mass spectrometry of large biomolecules and nanoparticles. As nanoscale objects land on NEMS sensor one by one, they induce resolvable shifts in the resonance frequency of the sensor proportional to their weight. The operational regime of NEMS sensors is often limited by the onset-of-nonlinearity, beyond which the highly sensitive schemes based on frequency tracking by phase-locked loops cannot be readily used. Here, we develop a measurement architecture to operate at the nonlinear regime and measure frequency shifts induced by analytes in a rapid and sensitive manner. We used this architecture to individually characterize the mass of gold nanoparticles and verified the results by performing independent measurements of the same nanoparticles based on linear mass sensing. Once the feasibility of the technique is established, we have obtained the mass spectrum of a 20 nm gold nanoparticle sample by individually recording about five hundred single particle events using two modes working sequentially in the nonlinear regime. The technique obtained here can be used for thin nanomechanical structures which possess a limited dynamic range.

Nanoelectromechanical Systems (NEMS) offer important advantages for mass sensing applications. In the last decade, the detection of single proteins,[1] mass resolution at the atomic[2-4] and near single-Dalton level,[5] mass spectrometry at the single-protein level,[6] and mass measurements of neutral species[7] have all been demonstrated. It was further shown that the information about the spatial distribution of analytes can be obtained by using multiple modes.[8-9] More recently, the efficient transportation to and characterization of virus capsids by NEMS sensors[10] have been reported. These advances suggest that NEMS based mass spectrometry offers a competitive alternative to conventional mass spectrometry especially for analytes with molecular weight above the Mega-Dalton range.

Two aspects of NEMS devices are critical for high mass sensitivity: device miniaturization and the precise detection of the resonance frequency of the sensing structure. The former provides a strong, fourth-power scaling for the responsivity of the sensor[11] while the latter enables very small perturbations to be detected. However, certain limits are faced when optimizing both aspects. For example, if the device width is decreased to increase sensitivity —while keeping the device length constant for transduction efficiency— the linear regime of operation shrinks.[12] For certain geometries, even thermal fluctuations are sufficient to drive the resonator into the nonlinear regime.[12] This decrease in the dynamic range prohibits the use of such device architectures since the common practice in the field has been to keep the devices on resonance at the linear regime. To alleviate this limitation, many studies have sought to increase the linear dynamic range by suppressing nonlinearity.[13-15] On the frequency detection aspect, the trend in the field has been to increase the drive power to decrease frequency noise and thereby increase the mass resolution. Although amplitude noise gets converted to phase noise in the nonlinear regime and environmental-induced frequency fluctuations increase with the increasing drive levels,[16] sensing in the nonlinear regime provides additional handles on the system. For instance, by fine tuning the feedback parameters, reducing the total phase noise of the sensor is still possible.[17-18] Moreover, as smaller sensors generate smaller signals, the ability to operate beyond the linear regime becomes critical to obtain a decent Signal-to-Noise Ratio (SNR). For these reasons, operation at the nonlinear regime holds great promise for sensing applications.

While the autonomous oscillator architecture offers excellent controllability[17,19-20], it is not always possible to build an oscillator circuit with nanomechanical devices since the signal-to-background ratio is usually small especially for smaller devices — making it difficult to satisfy Barkhausen condition only at the mechanical resonance frequency. Many of the work in the past used the open-loop response of nonlinear resonators,[21-24] including a recent technique for accurate characterization of frequency fluctuations in the nonlinear regime.[25] However, continuously sweeping the frequency in the open-loop cannot be applied effectively to the sensing of abrupt changes induced by single analytes for two reasons. First, open-loop technique requires judicious re-setting of the sweep parameters every time after a particle adsorption. Second, each frequency sweep needs to comprise many data points for sufficient precision which implies long sweep times: as such, the effective frequency noise increase due to long-term drift effects. Another sweep based technique[24] has utilized an extended frequency span and fast sweep parameters to calculate the particle-induced frequency shifts from the change in the amplitude response; however, since this technique is not adaptive, the accumulation of analytes would eventually shift the device parameters outside the sweep region. At the MEMS scale, the bifurcation sweep techniques near the amplitude jumps are reported[26-27] and the detection sensitivity in the nonlinear regime is shown to be advantageous over the linear regime for mass detection of gas molecules.[28] Essentially, the techniques developed so far have not been designed to track the frequency at the nonlinear bifurcation point in an adaptive, closed-loop manner and have not been used in single entity (nanoparticle/molecule) sensing.

The main ineffectiveness for the frequency tracking in the nonlinear regime comes from the lack of a powerful and robust method like phase locked loop (PLL) that is used for linear resonators. Although PLLs can conveniently track resonance frequency in the linear regime, the sharp phase transition and bistable response of nonlinear resonators (Figure 1b) prevent locking to a single phase at the resonance. Therefore, the nonlinear regime is generally avoided for mass sensing applications especially since performing a PLL does not look feasible in this regime.

Apart from the aforementioned issues in the closed-loop implementations, the sensing applications of nonlinear resonators have so far focused on chemical sensing in the gas phase. Here, we have performed mass and position sensing of single nanoparticles with the first two resonance modes by developing a robust technique to track the resonance frequencies in the nonlinear regime. We have achieved $10^{-6}$ Allan Deviation at about 1 second response time and collected about 500 single nanoparticle events and obtained the mass spectra of a 20 nm gold nanoparticle sample.

The device we used in the experiments is a 20 micron long, 320 nm wide, and 100 nm thick SiN device reported earlier.[29] In its linear regime, the phase response shows a sharp yet smooth transition (Figure 1a) which can be used as the reference target of a PLL circuit. When it is driven to the nonlinear regime though, hysteresis emerges and two different branches are observed depending on the sweep direction. More importantly, two sharp transitions (with theoretically infinite slope) in the phase are observed. The transition at the higher frequency is denoted as $f_{up}$ when sweeping from left to right and the one at the lower frequency is denoted as $f_{down}$ when sweeping from right to left. As shown before[25], these transition frequencies are related to the resonance frequency and effective mass of the structure. Therefore, continuously tracking either of these frequencies can be used to detect single particles landing on the structure. However, an architecture based on sweeping the frequency with an open loop configuration results in a slow response time and the corresponding Allan Deviation degrades due to long-term drift effects. On the other hand, building a feedback loop is very challenging due to the infinite slope at these transition frequencies.

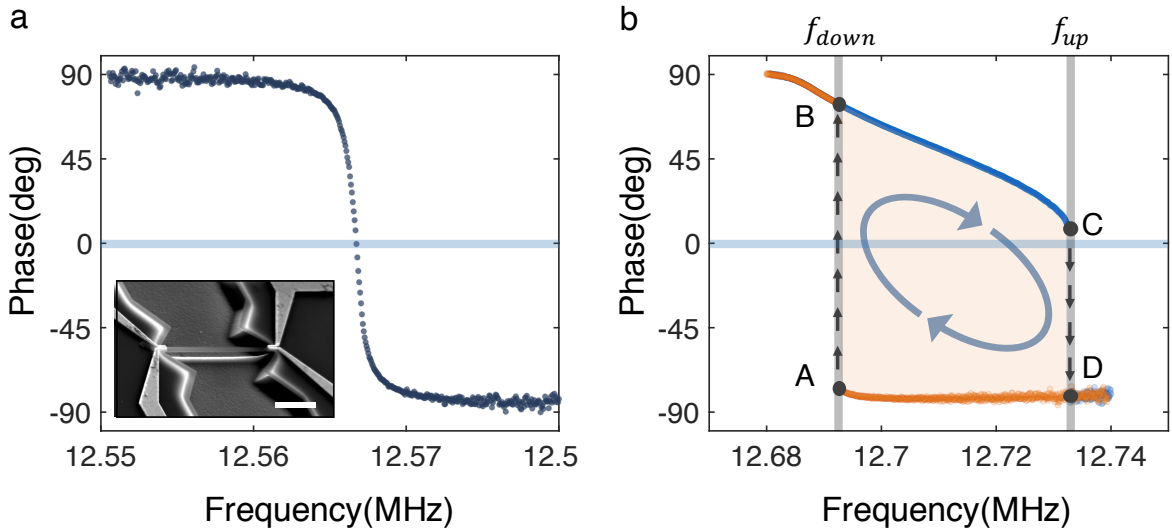

**Figure 1.** Linear and nonlinear response of the NEMS resonator (a) In the linear regime, a sharp phase transition is observed with the quality factor of 12000. 0° crossing in phase can be used as a reference target for PLL. The inset shows the SEM image of a typical doubly-clamped beam resonator used in the experiments. The scale bar is 3 $\mu m$. (b) Nonlinear phase jumps are observed depending on the sweep direction (blue and orange data points for the sweep from left to right and from right to left respectively). $f_{up}$ and $f_{down}$ frequencies are defined at the boundaries of the hysteresis window. In the colored area, the resonator shows a bistable response. 0° phase cannot be locked with the PLL as it is in the unstable region of the nonlinear response.

To overcome this problem, we aimed to keep the sensor trapped inside the hysteresis window of the phase response (Figure 1b) rather than locking to a single phase. Boundaries of the hysteresis window are defined as the points where the sharp frequency jumps ($f_{up}$ and $f_{down}$) occur. At the upper boundary, phase jumps from point C to D on the curve in Figure 1b, and at the lower boundary phase jumps from point A to B. Therefore, any controller which tries to be locked to a target phase at the sensitive jump frequencies cannot succeed since noise would push the PLL out of the operation point. However, the jump frequencies can be used for keeping the system circulating inside the hysteresis window.

To understand the method, we consider the nonlinear resonator with the phase response given in Figure 1b. We assume that there is a feedback controller with the target phase at 0 degrees: none of the stable branches shown in Figure 1b contains this point: indeed this point lies only on the unstable branch of the resonator (not shown in Figure 1b), hence it is not accessible within this measurement architecture. Therefore, the controller cannot keep the system locked at 0 degrees: however, a different dynamic emerges under these conditions. Whenever the phase has a positive value, the controller will increase the driving frequency; and whenever, the phase has a negative value, the controller will decrease the driving frequency. For this reason, starting from a random point, the controller will first push the system to the boundary of the window (i.e. until when the drive frequency is either $f_{up}$ or $f_{down}$, depending on the initial condition). After passing through either of the jump frequencies ($f_{up}$ or $f_{down}$), the sign of the phase flips, therefore *the controller action reverses automatically* and the system now starts traversing the other branch in the opposite direction. In effect, the system continuously circulates within the hysteresis window (Figure 1b), automatically tracing the boundary defined by the two jump frequencies. Although the control system is similar to the PLL, no phase is locked in this system, therefore we cite the proposed method as a trajectory-locked loop (TLL) for the convenience.

TLL can be used to analyze frequency fluctuations of the nonlinear bifurcation points.[25] During one cycle of TLL, it is possible to extract both the values of $f_{up}$ $and$ $f_{down}$ by looking at the sign of the derivative of the frequency with respect to time (Figure 2a-b). As it is demonstrated in Figure 2b, $f_{up}$ can be identified as the point where the derivative changes sign from positive to negative and vice versa for $f_{down}$. Figure 2c shows TLL operation in both phase and frequency domains. The phase response of the nonlinear resonator passes through a similar trajectory over time and keeps the system inside the hysteresis window as it can be seen at the projection of the frequency-phase plane. Projection of the data onto the phase-time plane clearly illustrates the phase jumps at the boundaries of the bistable regime.

The speed and precision to estimate the bifurcation points in one TLL cycle depend on the controller architecture, which is one of the main characteristics that distinguishes TLLs from PLLs. Even though the integration controller over phase error is used to prevent offset in PLLs, it causes overshoots for capturing bifurcation points for TLLs. For instance, if we consider the lower bound of one TLL cycle where the phase is negative and the controller steers the frequency from right to left, the error will accumulate with the integration controller action and just as the bifurcation point is passed, the accumulated error will still try to keep the same controller direction (whereas the direction should change). The same situation holds for the upper bound, consequently we found that the integrative controller, an essential part of PLLs, causes overshoots on the frequency measurements for TLLs. In order to increase the precision of the controller while detecting the bifurcation frequencies, we used the threshold phase values at the boundaries of the unstable regime (point C and point A in Figure 1b) for the error calculation. The controller is designed to adjust the frequency changes proportional to its distance from the boundaries of the hysteresis window, as evident in Figure 2b. In other words, when the phase of the resonator comes closer to one of the bifurcation thresholds, frequency steps between each sweep are decreased in order to reduce the offset error. Moreover, a larger step size while the resonator is away from the jump points increases the speed of the operation. In this way, accurate and fast measurements for $f_{up}$ and $f_{down}$ are achieved. More details on controller architecture for TLL are provided in SI Section

2. In Figure 2a, these two frequencies are measured with an averaging time 600ms, which is sufficient short to avoid drift effects.

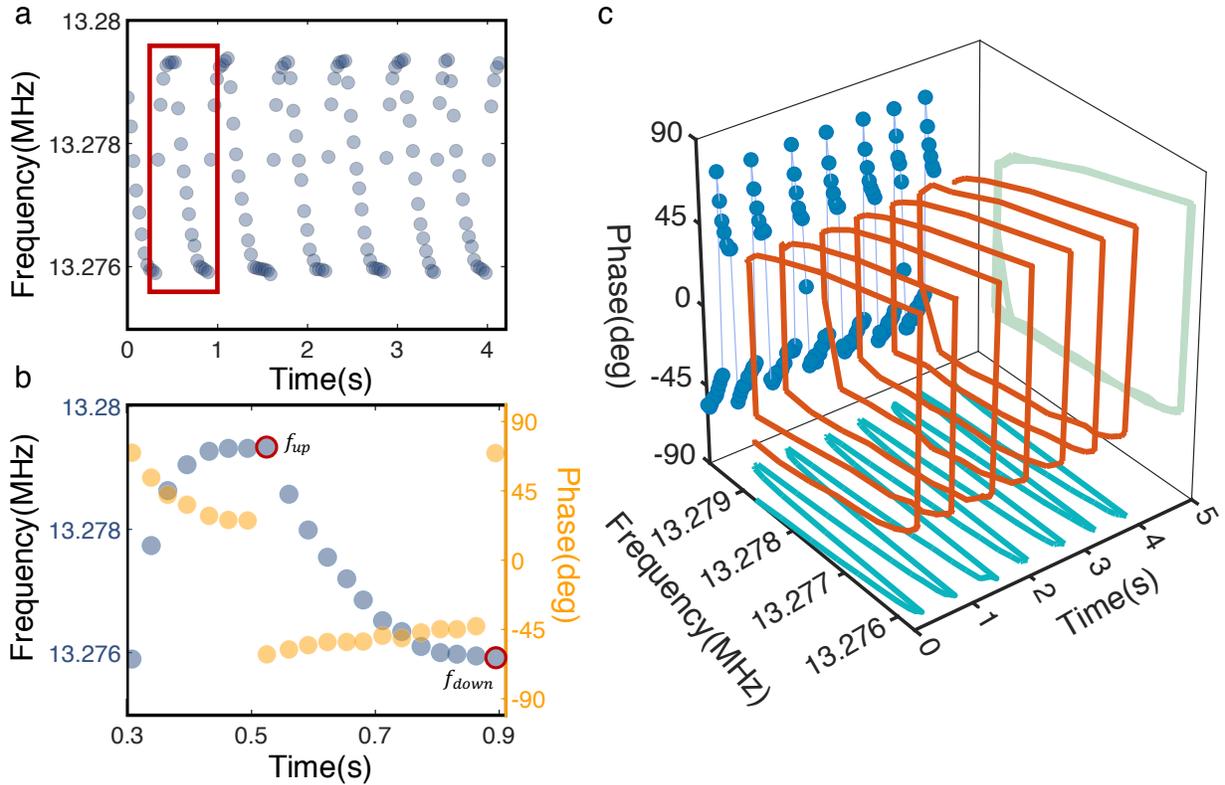

**Figure 2.** Trajectory Locked Loop (TLL). (a) The controller is adjusted for highly sensitive measurements of bifurcation frequencies. Circled data points show the frequency sweep steps which becomes denser while getting closer to the boundaries. (b) The one full cycle indicated with a red rectangle in (a) is maximized and plotted with the corresponding phase response. It is possible to extract $f_{up}$ and $f_{down}$ at the points where the phase jumps with a near infinite slope as they are indicated with red circles. For the given case, notice that one cycle approximately takes 0.6 seconds which can is comparable with many of the PLLs used in this field. (c) The projection of the TLL over time is displayed. It can be clearly seen that TLL holds the nonlinear resonator inside the hysteresis window.

After showing that TLL can track the bifurcation frequencies, we further used it to characterize the frequency fluctuations of the first two modes by calculating Allan deviations in the nonlinear regime. The ability to operate in nonlinear regime provides a wide-range of drive powers to be applied for the actuation of the resonator. Therefore, we calculated Allan deviations of the first two modes at different power levels in order to find the appropriate drive for the mass-position sensing of 20 nm gold nanoparticles (SI Section 3, and SI Figure 3-4). The Allan deviations at the chosen level of the nonlinear drive for the lower bifurcation frequencies ($f_{down}$) of the first two modes are determined as $1.5 \ 10^{-6}$ and $1.45 \ 10^{-6}$, respectively, for the chosen TLL response time of 2s, which corresponds to a mass resolution of $\sim 1.5$ MDa.

As we want to use the nonlinear resonators for single nanoparticle detection, we need to measure the analyte-induced frequency shifts of the first two modes.[6, 30-31] However, exciting the two modes simultaneously poses a challenge, since intermodal coupling[29, 32-35] —which may interfere with analyte-induced frequency changes— becomes more pronounced as the mode amplitudes reach nonlinear regime. Thus, extra care is needed for two-mode sensing with nonlinear resonators. To avoid the interference of coupling effects between the modes, we use TLL sequentially as it is demonstrated in Figure 3a. In this method, as one full-cycle is completed inside the hysteresis window for the first mode (meaning that $f_{1_{up}}$ and $f_{1_{down}}$ are detected), another cycle starts for the second mode (so that $f_{2_{up}}$ and $f_{2_{down}}$ are detected next). We note that, to avoid time delays due to power switching between modes during the sequential TLL operation, the inactive mode continues to be driven at the constant frequency which is close to the end frequency of its previous TLL cycle. As the cycle finishes at the lower bound, the inactive mode stays on the low-amplitude branch while the active mode circulates inside its TLL cycle. Thus, the intermodal coupling throughout the measurements is minimized.

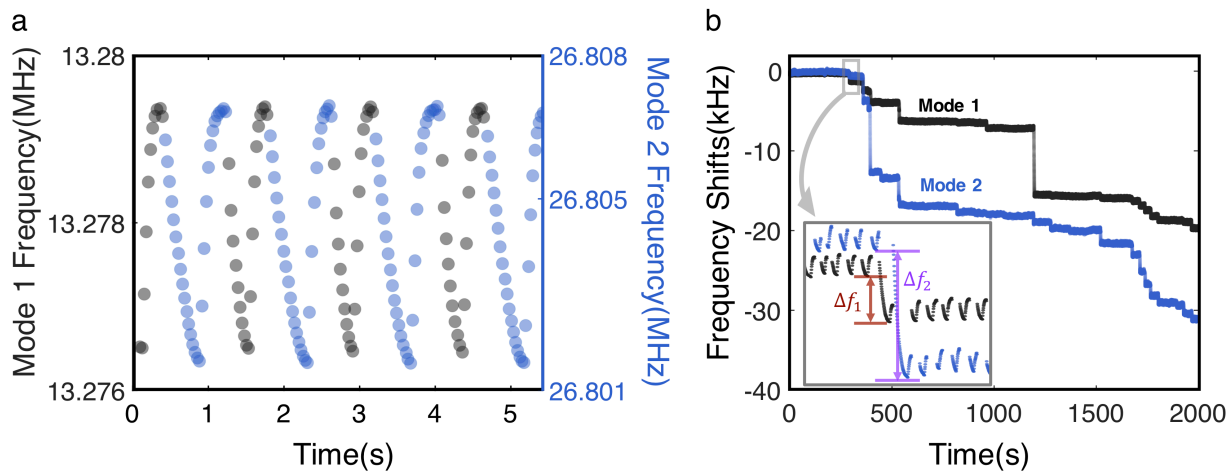

**Figure 3.** GNP sensing with Sequential TLL. (a) First and second resonance modes are tracked by TLL sequentially in order to minimize coupling between two nonlinear modes. After one mode completes the one full cycle inside the hysteresis window, the other mode is driven by TLL. (b) The frequency shifts due to individual GNP deposition. Each GNP adsorption causes frequency shifts in both modes, which emerge as sudden shifts in the trajectories (inset).

The frequency shift caused by an analyte is expected to be larger in the nonlinear regime than its counterpart in the linear regime.[25, 36] However, the normalized frequency shift (absolute frequency shift over the resonance frequency) due to an analyte is expected to be the same in both cases. In order to verify this equality, we have built a measurement system to sequentially switch between the linear and nonlinear operations, therefore the frequency shifts from the same nanoparticle can be directly compared with each other. As expected from the theory, the fractional frequency shifts measured by both techniques had resulted in essentially the same values within the measurement uncertainty (SI Section 4, and Figure S5). Once the feasibility of the technique is thus established, we have used our sensor to characterize a commercially available 20 nm Gold Nanoparticle solution (Sigma-Aldrich Product No: 741965). Mass spectrometry of the individual gold nanoparticles is performed by using matrix-assisted laser desorption ionization (MALDI) method.[6] Figure 3b demonstrates the snapshot of the two-mode sequential TLL data during the MALDI deposition of gold nanoparticles. Although it looks very similar to the PLL data at first glance, inset discloses the circular trajectories which are special to TLLs.

For the validation of the proposed method for mass sensing, we used normalized frequency shift of the lower bifurcation point ($f_{down}$) since it has a smaller noise level (as expected from[25] and also shown in the SI). As the normalized frequency shifts in nonlinear regime are shown to be same with the linear sensing, the earlier formulation for converting two-mode frequency shifts to the mass and position values[6,30] may directly be reused in this case. In Figure 4, we present mass spectra for 500 gold nanoparticles with a nominal diameter of 20 nm (12% dispersion in size) obtained by projecting the individual mass distributions onto the mass and diameter planes. For the mass measurements, the peak with maximum probability density is found at 57.25 MDa which corresponds to a diameter of 20.84 nm. The mean value of the gold nanoparticle sample is measured as 26.14 nm with a standard deviation of 5.73 nm. As evident from the mass spectrum, as well as the SEM image shown in Figure S6, a small portion of the nanoparticles have coalesced either in the solution or on the MALDI plate. For this reason, the mass distribution shows a fat tail with a few outliers on the high-end of the spectrum with a combined effect of shifting the statistical average of the mass distribution to a value higher than the nominal value.

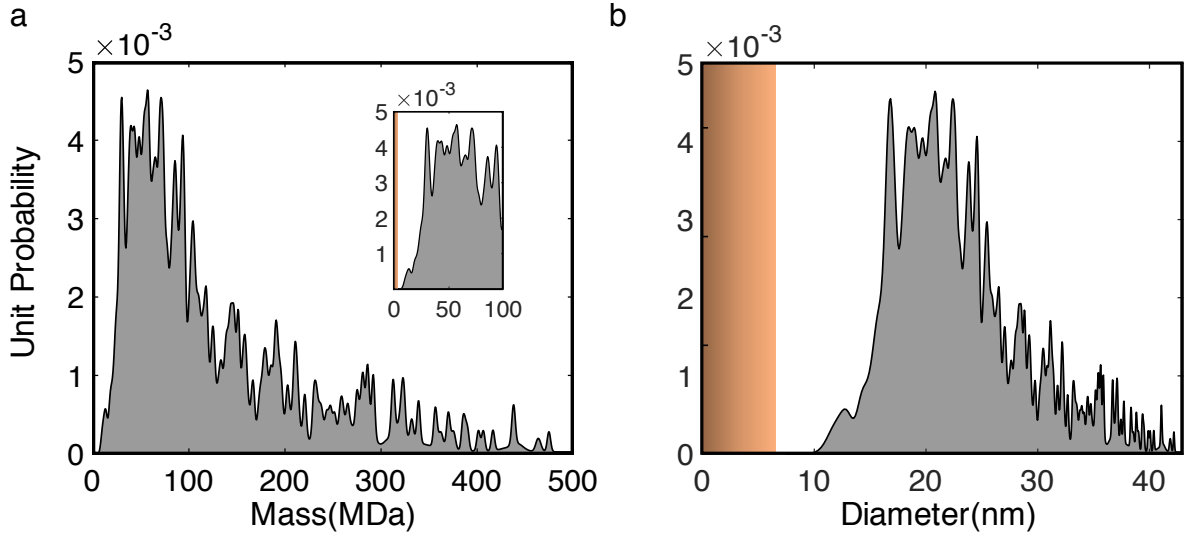

**Figure 4.** (a) Mass spectrometry of 20 nm gold nanoparticles. (b) Diameter is calculated with the bulk density of gold ($\rho_{Au} = 19.3 \frac{g}{cm^3}$). Orange-shaded regions illustrate the detection criterion due to frequency fluctuations of the nonlinear resonator used in the experiments.

In this work, we present a rapid and accurate method with a feedback controller for trapping the nonlinear resonator alongside the regime of bistability (TLL). By this technique, we can conveniently and precisely track the bifurcation frequencies. Later, we use TLL to characterize the frequency instabilities at these bifurcation points for different averaging times by calculating the Allan deviations. Unlike the linear dynamic range, the nonlinear region does not restrict the amplitude of the resonance at critical amplitude, therefore we further analyze the frequency fluctuations at much larger amplitudes. After we adjust the control and nonlinearity parameters for the nonlinear resonator, we test the feasibility of TLL for the single particle sensing. Results validate that TLL shows a remarkable performance for determining the frequency shifts due to adsorbed particles, therefore can be used for mass spectrometry applications within the nonlinear regime of the NEMS resonators.

Nonlinearity in NEMS resonators has long been acknowledged as a resource, however, its usage in sensing applications has generally been avoided due to the complexity of feedback circuits. Whereas, PLL systems have provided a means for frequency tracking of linear resonators rapidly which was absent in the nonlinear region until now. We demonstrated that a feedback controller similar to PLL —but circulating between two points rather than locking onto a point— can be used for reliable frequency tracking in nonlinear response.

NEMS resonators have been shrinking in size— a trend which will result in a reduced, and even nonexistent, linear dynamic range. The performance of *inherently* nonlinear resonators for single particle and molecule detection is still unknown due to the lack of robust techniques for enabling nonlinear frequency tracking. Our method can be deployed on such systems rather effortlessly and help us further analyze the potential of nonlinearity in NEMS sensors. Moreover, the applicability of the method is not limited with the mechanical resonators as it can be adjusted to any resonator with the nonlinear response.

ASSOCIATED CONTENT

**Supporting Information.** Experimental setup, details regarding controller of TLL, measured frequency fluctuations for the two modes used for sensing, analysis on particle induced frequency shifts in nonlinear regime, figures of the resonator used in the experiments (PDF file).

AUTHOR INFORMATION

**Corresponding Author**

*E-mail: selimhanay@bilkent.edu.tr

**Present Addresses**

† The present address of EO and ABA is Boston University, Boston, MA.

**Author Contributions**

The manuscript was written through contributions of all authors. All authors have given approval to the final version of the manuscript. MY has conceived the idea: MY, MSH and AD have developed the idea further. MSH, ABA and CY designed the devices. CY, ABA, EO and MY fabricated the devices. EO, MY and ABA have built the experimental setup.

**Funding Sources**

This work was supported by the Scientific and Technological Research Council of Turkey (**TÜBİTAK**), Grant No: EEEAG-115E230. MSH acknowledges support from TÜBA and Bilim Akademisi.

REFERENCES

1. Naik, A. K.; Hanay, M. S.; Hiebert, W. K.; Feng, X. L.; Roukes, M. L., Towards single-molecule nanomechanical mass spectrometry. *Nature Nanotechnology* **2009,** *4* (7), 445-450.

2. Jensen, K.; Kim, K.; Zettl, A., An atomic-resolution nanomechanical mass sensor. *Nature nanotechnology* **2008,** *3* (9), 533.

3. Chiu, H.-Y.; Hung, P.; Postma, H. W. C.; Bockrath, M., Atomic-scale mass sensing using carbon nanotube resonators. *Nano letters* **2008,** *8* (12), 4342-4346.

4. Lassagne, B.; Garcia-Sanchez, D.; Aguasca, A.; Bachtold, A., Ultrasensitive mass sensing with a nanotube electromechanical resonator. *Nano letters* **2008,** *8* (11), 3735-3738.

5. Chaste, J.; Eichler, A.; Moser, J.; Ceballos, G.; Rurali, R.; Bachtold, A., A nanomechanical mass sensor with yoctogram resolution. *Nature nanotechnology* **2012,** *7* (5), 301.

6. Hanay, M. S.; Kelber, S.; Naik, A. K.; Chi, D.; Hentz, S.; Bullard, E. C.; Colinet, E.; Duraffourg, L.; Roukes, M. L., Single-protein nanomechanical mass spectrometry in real time. *Nature Nanotechnology* **2012,** *7* (9), 602-608.

7. Sage, E.; Brenac, A.; Alava, T.; Morel, R.; Dupré, C.; Hanay, M. S.; Roukes, M. L.; Duraffourg, L.; Masselon, C.; Hentz, S., Neutral particle mass spectrometry with nanomechanical systems. *Nature communications* **2015,** *6*.

8. Hanay, M. S.; Kelber, S. I.; O'Connell, C. D.; Mulvaney, P.; Sader, J. E.; Roukes, M. L., Inertial imaging with nanomechanical systems. *Nature nanotechnology* **2015,** *10* (4), 339-344.

9. Sader, J. E.; Hanay, M. S.; Neumann, A. P.; Roukes, M. L., Mass spectrometry using nanomechanical systems: beyond the point-mass approximation. *Nano letters* **2018,** *18* (3), 1608-1614.

10. Dominguez-Medina, S.; Fostner, S.; Defoort, M.; Sansa, M.; Stark, A.-K.; Halim, M. A.; Vernhes, E.; Gely, M.; Jourdan, G.; Alava, T.; Boulanger, P.; Masselon, C.; Hentz, S., Neutral mass spectrometry of virus capsids above 100 megadaltons with nanomechanical resonators. *Science* **2018,** *362* (6417), 918-922.

11. Li, M.; Tang, H. X.; Roukes, M. L., Ultra-sensitive NEMS-based cantilevers for sensing, scanned probe and very high-frequency applications. *Nature nanotechnology* **2007,** *2* (2), 114-120.

12. Postma, H. C.; Kozinsky, I.; Husain, A.; Roukes, M., Dynamic range of nanotube-and nanowire-based electromechanical systems. *Appl Phys Lett* **2005,** *86* (22), 223105.

13. Kacem, N.; Arcamone, J.; Perez-Murano, F.; Hentz, S., Dynamic range enhancement of nonlinear nanomechanical resonant cantilevers for highly sensitive NEMS gas/mass sensor applications. *Journal of Micromechanics and Microengineering* **2010,** *20* (4), 045023.

14. Kozinsky, I.; Postma, H. C.; Bargatin, I.; Roukes, M., Tuning nonlinearity, dynamic range, and frequency of nanomechanical resonators. *Appl Phys Lett* **2006,** *88* (25), 253101.

15. Samanta, C.; Arora, N.; Naik, A., Tuning of geometric nonlinearity in ultrathin nanoelectromechanical systems. *Appl Phys Lett* **2018,** *113* (11), 113101.

16. Sansa, M.; Sage, E.; Bullard, E. C.; Gély, M.; Alava, T.; Colinet, E.; Naik, A. K.; Villanueva, L. G.; Duraffourg, L.; Roukes, M. L., Frequency fluctuations in silicon nanoresonators. *Nature nanotechnology* **2016,** *11* (6), 552-558.

17. Kenig, E.; Cross, M.; Villanueva, L.; Karabalin, R.; Matheny, M.; Lifshitz, R.; Roukes, M., Optimal operating points of oscillators using nonlinear resonators. *Physical Review E* **2012,** *86* (5), 056207.

18. Yurke, B.; Greywall, D.; Pargellis, A.; Busch, P., Theory of amplifier-noise evasion in an oscillator employing a nonlinear resonator. *Physical Review A* **1995,** *51* (5), 4211.

19. Chen, C.; Zanette, D. H.; Guest, J. R.; Czaplewski, D. A.; López, D., Self-sustained micromechanical oscillator with linear feedback. *Physical review letters* **2016,** *117* (1), 017203.

20. Demir, A.; Hanay, M. S., Numerical Analysis of Multidomain Systems: Coupled Nonlinear PDEs and DAEs With Noise. *IEEE Transactions on Computer-Aided Design of Integrated Circuits and Systems* **2018,** *37* (7), 1445-1458.

21. Kumar, V.; Boley, J. W.; Yang, Y.; Ekowaluyo, H.; Miller, J. K.; Chiu, G. T.-C.; Rhoads, J. F., Bifurcation-based mass sensing using piezoelectrically-actuated microcantilevers. *Appl Phys Lett* **2011,** *98* (15), 153510.

22. Harne, R.; Wang, K., Robust sensing methodology for detecting change with bistable circuitry dynamics tailoring. *Appl Phys Lett* **2013,** *102* (20), 203506.

23. Venstra, W. J.; Capener, M. J.; Elliott, S. R., Nanomechanical gas sensing with nonlinear resonant cantilevers. *Nanotechnology* **2014,** *25* (42), 425501.

24. Sansa, M.; Nguyen, V. N.; Baguet, S.; Lamarque, C.-H.; Dufour, R.; Hentz, S. In *Real time sensing in the non linear regime of nems resonators*, IEEE 29th International Conference on Micro Electro Mechanical Systems (MEMS), 2016.

25. Maillet, O.; Zhou, X.; Gazizulin, R. R.; Ilic, R.; Parpia, J. M.; Bourgeois, O.; Fefferman, A. D.; Collin, E., Measuring frequency fluctuations in nonlinear nanomechanical resonators. *ACS nano* **2018**.

26. Requa, M. V.; Turner, K. L., Precise frequency estimation in a microelectromechanical parametric resonator. *Appl Phys Lett* **2007,** *90* (17), 173508.

27. Burgner, C.; Miller, N.; Shaw, S.; Turner, K. In *Parameter sweep strategies for sensing using bifurcations in MEMS*, Solid-State Sensor, Actuator, and Microsystems Workshop, Hilton Head Workshop, 2010.

28. Yie, Z.; Zielke, M. A.; Burgner, C. B.; Turner, K. L., Comparison of parametric and linear mass detection in the presence of detection noise. *Journal of Micromechanics and Microengineering* **2011,** *21* (2), 025027.

29. Arı, A. B.; Karakan, M. Ç.; Yanık, C.; Kaya, İ. İ.; Hanay, M. S., Intermodal Coupling as a Probe for Detecting Nanomechanical Modes. *Physical Review Applied* **2018,** *9* (3), 034024.

30. Dohn, S.; Svendsen, W.; Boisen, A.; Hansen, O., Mass and position determination of attached particles on cantilever based mass sensors. *Rev Sci Instrum* **2007,** *78* (10), 103303.

31. Schmid, S.; Dohn, S.; Boisen, A., Real-time particle mass spectrometry based on resonant micro strings. *Sensors* **2010,** *10* (9), 8092-8100.

32. Matheny, M.; Villanueva, L.; Karabalin, R.; Sader, J. E.; Roukes, M., Nonlinear mode-coupling in nanomechanical systems. *Nano letters* **2013,** *13* (4), 1622-1626.

33. Westra, H.; Poot, M.; Van Der Zant, H.; Venstra, W., Nonlinear modal interactions in clamped-clamped mechanical resonators. *Physical review letters* **2010,** *105* (11), 117205.

34. Eichler, A.; del Álamo Ruiz, M.; Plaza, J.; Bachtold, A., Strong coupling between mechanical modes in a nanotube resonator. *Physical review letters* **2012,** *109* (2), 025503.

35. Truitt, P.; Hertzberg, J.; Altunkaya, E.; Schwab, K., Linear and nonlinear coupling between transverse modes of a nanomechanical resonator. *Journal of Applied Physics* **2013,** *114* (11), 114307.

36. Dai, M. D.; Eom, K.; Kim, C.-W., Nanomechanical mass detection using nonlinear oscillations. *Appl Phys Lett* **2009,** *95* (20), 203104.